\documentclass{ws-ijmpcs}

\begin{document}

\markboth{L. Hilico \& al.}
{Cooling antihydrogen ions for the free-fall experiment GBAR}

%
\catchline{}{}{}{}{}
%

\title{Preparing single ultra-cold antihydrogen atoms for the free-fall in GBAR}

\author{LAURENT HILICO, JEAN-PHILIPPE KARR, ALBANE DOUILLET}

\address{D\'epartement de Physique, Universit\'e d'Evry Val d'Essonne\\
Rue du p\`ere Andr\'e Jarlan\\ Evry 91025, France\\
Laboratoire Kastler Brossel UMR 8552, UPMC, CNRS, ENS, C 74, Universit\'e Pierre et Marie Curie\\
4 place Jussieu, Paris, 75252, France\\
Laurent.hilico@lkb.upmc.fr, jean-philippe.karr@lkb.upmc.fr, albane.douillet@lkb.upmc.fr}

\author{PAUL INDELICATO}

\address{Laboratoire Kastler Brossel UMR 8552, UPMC, CNRS, ENS, C 74, Universit\'e Pierre et Marie Curie\\
4 place Jussieu, Paris, 75252, France\\
paul.indelicato@lkb.upmc.fr}

\author{SEBASTIAN WOLF and FERDINAND SCHMIDT KALER}
\address{QUANTUM, Institut f\"ur Physik, Universit\"at Mainz, D-55128 Mainz, Germany\\
fsk@uni-mainz.de, wolfs@uni-mainz.de}

\maketitle

\begin{history}
\received{Day Month Year}
\revised{Day Month Year}
\end{history}

\begin{abstract}
We discuss an experimental approach allowing to prepare antihydrogen atoms for the GBAR experiment.
We study the feasibility of all necessary experimental steps: The capture of incoming $\bar{\rm H}^+$ ions at keV energies in a deep linear RF trap, sympathetic cooling by laser cooled Be$^+$ ions, transfer to a miniaturized trap and Raman sideband cooling of an ion pair to the motional ground state, and further reducing the momentum of the wavepacket by adiabatic opening of the trap.
For each step, we point out the experimental challenges and discuss the efficiency and characteristic times, showing that capture and cooling are possible within a few seconds.
\end{abstract}

\ccode{PACS numbers:37.10.Rs,36.10.-k,04.80.Cc,36.10.Gv}

\section{Introduction}
The GBAR project aims at measuring the earth gravity acceleration hereafter denoted $\bar{g}$ felt by an antihydrogen
atom $\bar{\rm H}$, using a free-fall technique at first\cite{Walz2004,Perez,SPSC342} and possibly spectroscopy of $\bar{\rm H}$
gravitational states in the future\cite{Voronin2012,Dufour2013}. Other collaborations AEGIS\cite{Aegis}, ATHENA-ALPHA\cite{Alpha}, ATRAP\cite{Atrap} pursue the same goal using different methods.
The specificity of the GBAR project is to prepare a single antihydrogen atom at a temperature of the order of 10~$\mu$K, to obtain a sub-percent accuracy\cite{Dufour2014} on $\bar{g}$.
After a brief explanation of the scheme proposed for the GBAR experiment, we discuss in detail the trapping and sympathetic cooling of antihydrogen ions.

In the GBAR experimental scheme\cite{Walz2004,SPSC342}, neutral antihydrogen is prepared by photodetachment of the excess positron of a sympathetically
cooled $\bar{\rm H}^+$ trapped ion. The photodetachment is performed using a pulsed laser. The start time of the
free-fall is the photodetachment pulse time. The stop time corresponds to the annihilation of the $\bar{\rm H}$ on the
detection plate. Obviously, reducing the velocity spread of $\bar{\rm H}^+$ atoms is indispensable for determining $\bar{g}$ with high precision.

The $\bar{\rm H}^+$ ions are produced in two steps in a collision cell by sending keV antiprotons from the ELENA ring on a room temperature
positronium cloud. The first step produces antihydrogen atoms and the second one the $\bar{\rm H}^+$ ions
following the reactions:
\begin{equation}
\bar{p}+Ps \rightarrow \bar{\rm H}+e^-\hspace{0.5cm}{\rm and}\hspace{0.5cm}\bar{\rm H}+Ps \rightarrow \bar{\rm H}^+ + e^-.
\end{equation}
The reaction cross sections have been evaluated by P. Comini et al.\cite{Comini2013}, predicting that bunches of a few $\bar{\rm H}^+$ can be produced using state-of-the-art Ps sources.
Since positronium is much lighter than $\bar{\rm p}$, the $\bar{\rm H}^+$ energy distribution is linked to that of the
$\bar{\rm p}$ bunch produced by the ELENA ring\cite{Elena} (whose expected characteristics are 100~keV mean energy and a 4$\pi$~mm~mrad emittance). The $\bar{\rm p}$ bunch is decelerated to an energy of a few keV using a drift tube. The kinetic energy spread of the $\bar{\rm p}$ bunch and hence of the
$\bar{\rm H}^+$ ion bunch is about 300~eV corresponding to a temperature of 2.3$\times 10^{+6}$~K.

The relative resolution on $\bar{g}$ that can be obtained measuring the free-fall time of a single particle is given by\cite{Dufour2014}
\begin{equation}
\frac{\Delta\bar{g}}{\bar{g}}=2\sqrt{\left(\frac{\Delta\zeta}{2H}\right)^2+\left(\frac{\Delta v}{\sqrt{2\bar{g}H}}\right)^2}\label{eq_deltagsurg}
\end{equation}
where $\Delta\zeta$ and $\Delta v$ are the position and velocity dispersions in the vertical direction. In the case of a quantum particle at the Heisenberg limit, $\Delta\zeta$ and $\Delta v$ are linked by the uncertainty relation $m\Delta v\Delta\zeta=\hbar/2$ where $m$ is the $\bar{\rm H}$ inertial mass. An optimum resolution $(\Delta\bar{g}/\bar{g})_{opt}=2^{1/4}\hbar^{1/2}m^{-1/2}\bar{g}^{-1/4}H^{-3/4}$ is obtained for $\Delta v_{opt}=2^{-3/4}\hbar^{1/2}m^{-1/2}\bar{g}^{1/4}H^{-1/4}$ leading to
$\Delta v_{opt}=$~2.6~mm/s and $(\Delta\bar{g}/\bar{g})_{opt}=1.7\times10^{-4}$
assuming $\bar{g}=g$ and $H=1$~m.

The recoil due to the absorption of the 1.64~$\mu$m detachment photon is
$h/(m\lambda)=23$~cm/s and may be set in the horizontal plane so that its influence on the free-fall vanishes.
The recoil due to the
excess energy can be made small using threshold detachment. The associated energy is 0.3~m/s under the realistic assumption that the photon energy is 1~$\mu$eV above detachment threshold.
Thus photodetachment prevents reaching the optimal free-fall conditions even if the initial ion were perfectly motionless.  Our goal vertical velocity dispersion is $\Delta v\approx1$~m/s, Eq.~\ref{eq_deltagsurg} is then dominated by the second term leading to $\Delta\bar{g}/\bar{g}=\sqrt{2}\Delta v/\sqrt{\bar{g}H}$~=~0.4 per detected atom. A 1\% resolution can be obtained by averaging on 1600 events.

A transverse initial velocity of the
order of 1~m/s is also required to avoid a too large detection area for the $\bar{\rm H}^+$  annihilation plates.
Those velocities correspond to energies of the order
of 5.2~neV or 120~$\mu$K per degree of freedom.
The $\bar{\rm H}^+$ cooling challenge is to bridge a 10 to 11 orders of magnitude gap on the ion temperature,
going from the classical world of particle beam physics to the ultimate frontiers of quantum world.
Indeed, if one considers the ground state of a quantum harmonic oscillator of mass $m$~= 1~a.u. and angular frequency $\omega$,
the velocity spread is given by $\Delta v=\sqrt{\hbar\omega/2m}$. $\Delta v=1$~m/s leads to $\omega=2\pi\times
5$~MHz. This is the typical secular motion frequencies that are achieved in ion traps, showing that the GBAR
requirements can only be satisfied using ground state cooling techniques\cite{Leibfried2003}.

$\bar{\rm H}^+$, antimatter equivalent of H$^-$, is extremely fragile against collisions with regular matter such that buffer gas cooling is not possible. Moreover, $\bar{\rm H}^+$ is a single electronic level atom that
cannot be directly laser cooled. Hence, we propose to use sympathetic cooling by the lighest laser cooled ion: $^9$Be$^+$.

\section{Antihydrogen positive ion capture and Doppler cooling}\label{sec_capt_and_cooling}
Since only a few $\bar{\rm H}^+$ ions are expected in each bunch, a nearly 100\% capture efficiency is required.
To that end, the GBAR project will first use capture and Doppler laser cooling step in a mm scale RF linear trap before transferring a
single $\bar{\rm H}^+$ ion into a miniaturized trap (called precision trap in the following) to perform ground state cooling of a Be$^+$/$\bar{\rm H}^+$ ion pair.
In Sect.~\ref{sec_capt_and_cooling}, we discuss the capture and sympathetic cooling of a $\bar{H}^+$ ion in a big Be$^+$ crystal.
In Sect.~\ref{sec_precision_trap}, we discuss the separation of the cold $\bar{\rm H}^+$ ion and injection in the precision trap for ground state sideband sympathetic cooling before neutralization and $\bar{\rm H}$ release.

\subsection{$\bar{\rm H}^+$ ion capture}
\begin{figure}[pb]
\centerline{\includegraphics[width=7cm]{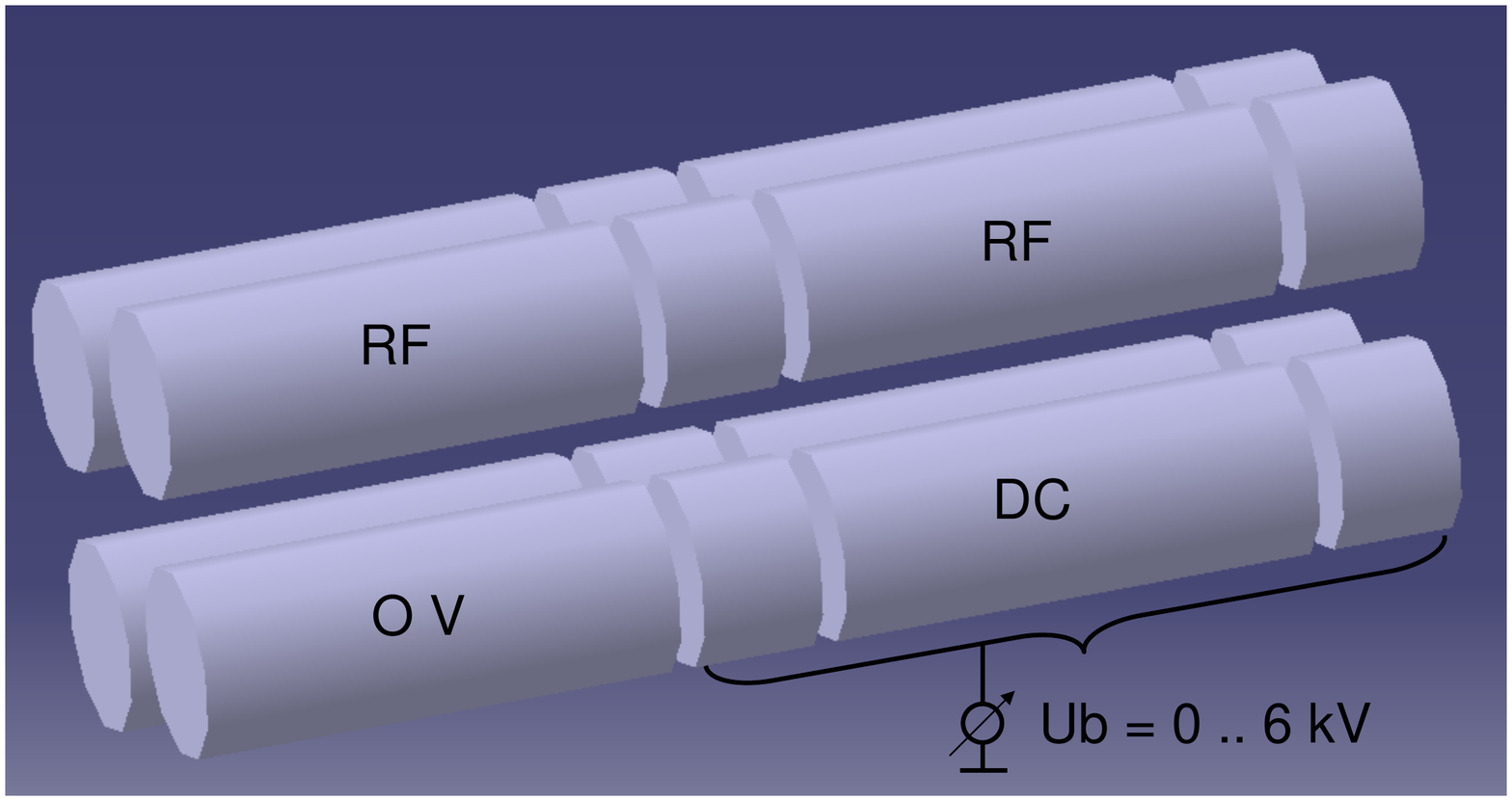}}
\caption{Quadrupole guide and biased linear RF Paul trap used to decelerate and trap the $\bar{\rm H}^+$ ion bunch. $L$=~30~mm.}\label{fig_capt_trap}
\end{figure}

The capture apparatus is depicted in Fig.~\ref{fig_capt_trap}.
It is made of a RF quadrupole guide and a biased segmented linear trap described below.
If the $\bar{p}$ are not precooled below 300~eV, the kinetic energy spread of the $\bar{\rm H}^+$ ion bunch is large, and requires a deep trap.
The trapping depth of a
linear trap is given by $U_{max}=qV_{RF}/8$ where q is the stability parameter and $V_{RF}$ the applied voltage. The $q$
parameter is inversely proportional to the ion mass and is typically chosen between 0.05 and 0.6 (larger values lead to important RF heating of the ions). In order to safely trap both Be$^+$ and $\bar{\rm H}^+$, $q$ must be chosen close to 0.45 for $\bar{\rm H}^+$
and 0.05 for Be$^+$.
We assume a trapping
depth of 20~eV for $\bar{\rm H}^+$ that is obtained using $V_{RF}=$~356~V (712~V peak to peak). The stability parameter of a linear trap is given by $q=2QV_{RF}/m\Omega^2r_0^2$ where $m$ and $Q$ are the ion mass and charge, $r_0$ is the inner
trap radius, and $\Omega$ the RF frequency. With $r_0=3.5$~mm, we get $\Omega=2\pi\times17.7$~MHz, i.e., standard trap parameters. Efficient capture of $\bar{\rm H}^+$ ions with a 300~eV kinetic energy spread is much more involved, requiring 10~800~V peak to peak at 68.5~MHz.
The use of RF traps with 2 drive frequencies\cite{Trypogeorgos2013} was envisaged but it requires 2-3 order of magnitude different mass-to-charge ratios.

\begin{figure}[pb]
\centerline{\includegraphics[width=4.7cm]{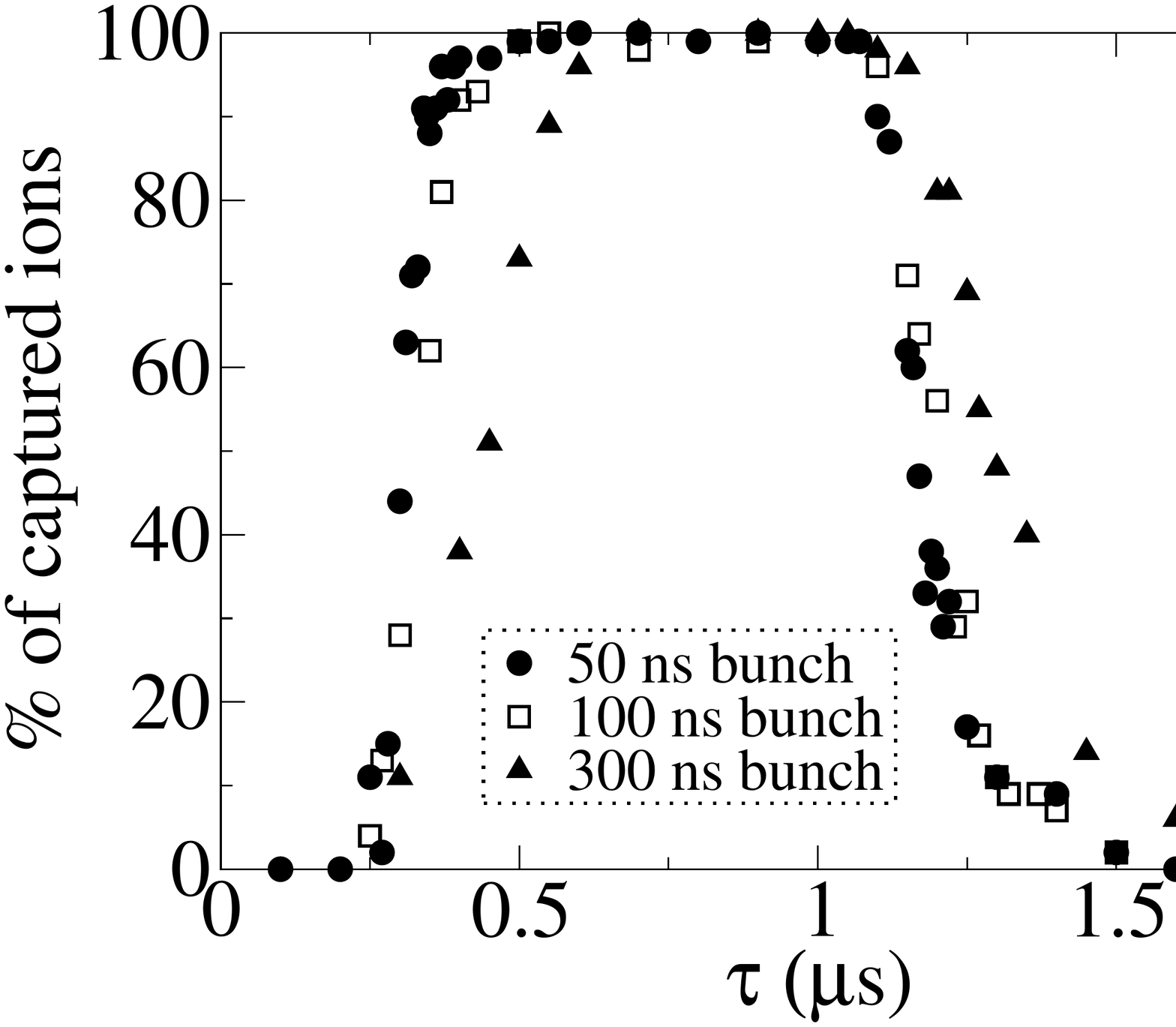}
\includegraphics[width=4.7cm]{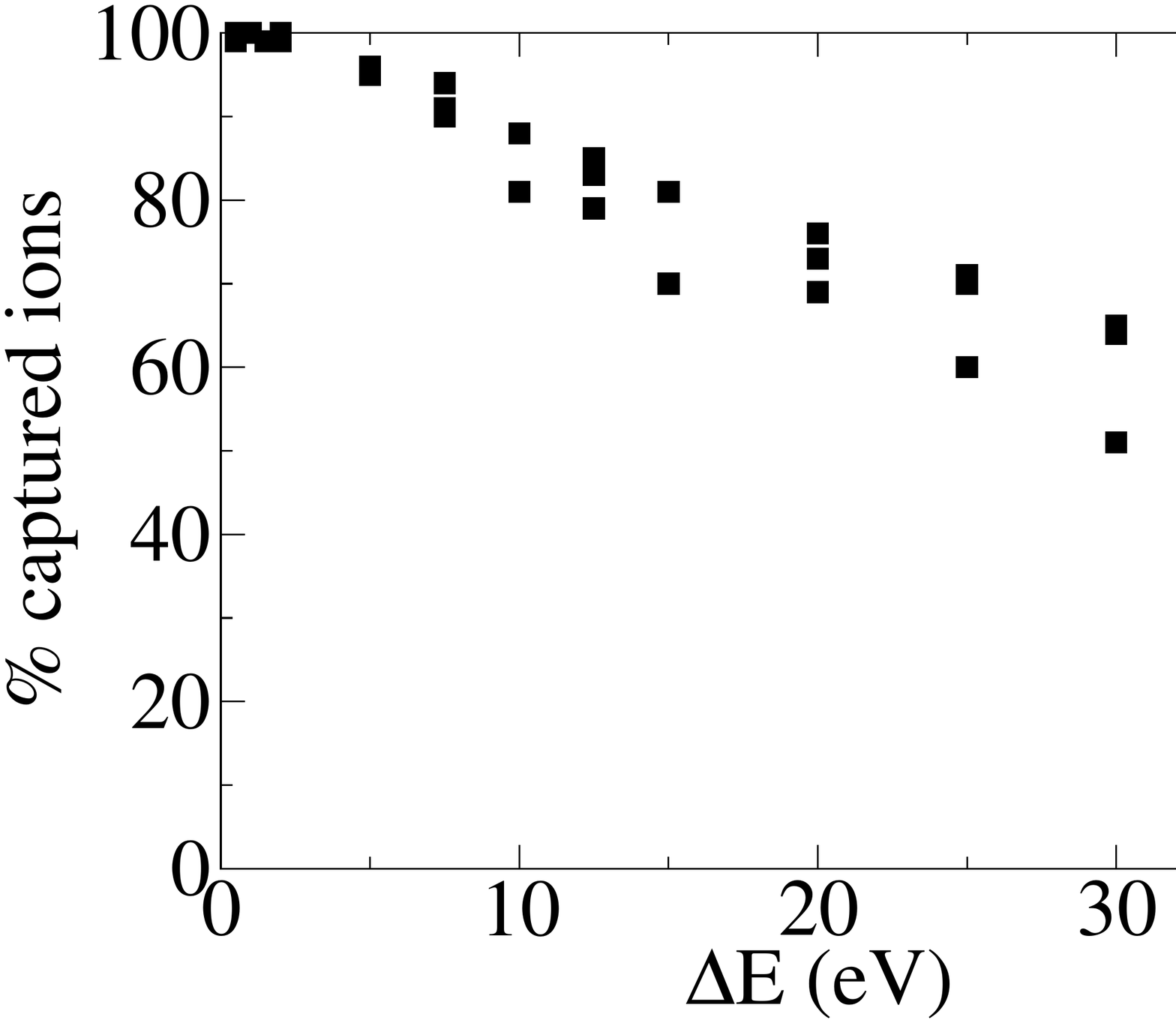}}
\vspace*{8pt}
\caption{(a): $\bar{\rm H}^+$ capture efficiency versus trap opening time $\tau$ for different ion bunch durations, for $\Delta E=1$~V.
(b): $\bar{\rm H}^+$ capture efficiency versus kinetic energy spread for $\tau=0.9\ \mu$s. Each point corresponds to a different simulation with 100 ions.}\label{fig_capt_eff_tau_ener}
\end{figure}

The incoming $\bar{\rm H}^+$ ion bunch has 1 to 6~keV kinetic energy. The ion bunch is decelerated by
biasing the linear trap by $U_b=$~980 to 5980~V. The trap input endcap voltage is lowered to $U_b$
for a short time $\tau$ for the $\bar{\rm H}^+$ ion bunch intake. Since the Be$^+$ ion motion is strongly
damped by the cooling laser, and $\tau$ is much shorter than the
axial trap secular period, the Be$^+$ ions don't have time to escape the trap.
Figure~\ref{fig_capt_eff_tau_ener}-a shows a simulation of the
capture efficiency (without Be$^+$) versus the time delay $\tau$. 100\% capture efficiency is predicted for a large range of $\tau$
for a small kinetic energy spread $\Delta E=1$~eV. Figure~\ref{fig_capt_eff_tau_ener}-b shows the capture efficiency for optimal intake time
$\tau$ versus the $\bar{\rm H}^+$ bunch kinetic energy spread $\Delta E$. The efficiency decreases with $\Delta E$, but
remains larger than 50\% for $\Delta E <$~25~eV.
This analysis shows that the initial kinetic energy spread of antiprotons from the ELENA source has to be reduced by at least one order of magnitude to allow for their efficient capture with reasonable trap parameters.

\subsection{Sympathetic Doppler cooling time}
Once captured, $\bar{\rm H}^+$ ions have a very high temperature, limited by the trap depth.
Next, they are sympathetically cooled by Coulomb interaction with
a large laser cooled Be$^+$ ion cloud.

Because of possible photodetachment by the cooling laser light (see Sect.~\ref{sec_photodetach}),
it is very important to evaluate the cooling time. Sympathetic cooling dynamics results from the competition
between the Coulomb repulsion and the trapping forces that take the ions together, and between
laser cooling and RF heating. For this reason, it can only be evaluated using ion dynamics numerical simulation
taking into account the exact time-dependent trapping forces responsible for micromotion and RF heating, and the exact Coulomb repulsion\cite{Lammerzahl2003}. Short time steps (sub-ns range) must be used to well represent the fast dynamics due to the RF field and secure calculation
convergence, since long simulation times are required to get sympathetic cooling evidence. The main numerical complication comes
from the evaluation of the full Coulomb interaction for a large number of ions.

The numerical simulations are done using a home built FORTRAN code to solve the Newton's equation of motion for $N_{lc}$ laser cooled Be$^+$ ions and $N_{sc}$ sympathetically cooled ions, whose masses were taken equal to 1 ($\bar{\rm H}^+$), 2 (H$_2^+$) and 3 (HD$^+$) in order to study the mass dependence of the cooling process.
The equations are integrated using either a fixed-step fourth-order Runge-Kutta method or the leap-frog (Verlet-velocity)\cite{Verlet1967} algorithm. The code takes into account the time-dependent RF trapping field and the axial harmonic trapping field of an ideal linear trap model given by the gradient of the potential
\begin{equation}
V(x,y,z,t)=(U_0+V_{\rm RF}cos(\Omega t))\frac{x^2-y^2}{2r_0^2}+m_i\omega_{i,z}^2 (z^2-(x^2+y^2)/2),\label{eq_RF_potential}
\end{equation}
where $V_{\rm RF}$ is the RF voltage, $\Omega$ the RF angular frequency, $r_0$ is the effective inner radius of the ion trap, $m_{(i)}$ is the mass of the considered ion and $\omega_{i,z}$ its axial oscillation frequency. For two different ionic species labeled $i$ and $j$, we have $m_i\omega_{i,z}^2=m_j\omega_{j,z}^2$.
The Coulomb force undergone by ion $i$ is given by
\begin{equation}
{\bf F}_{\rightarrow i}=\sum_{j\neq i}\frac{q_iq_j}{4\pi\epsilon_0}\frac{{\bf r}_i-{\bf r}_j}{r_{ij}^3}.
\end{equation}
The laser action is taken into account in terms of absorption, spontaneous and stimulated emission processes for a two-level atom in a Gaussian laser beam of waist $w_0$ and wave vector ${\bf k}$. At each time step and for each laser-cooled ion, depending on its internal state and position in the laser beam, the absorption and emission probabilities are evaluated in a quantum jump approach.
In case of absorption, stimulated or spontaneous emission, the ion velocity is changed by $\pm\hbar{\bf k}/m$ or by $\hbar k\boldsymbol{\kappa}/m$ where $\boldsymbol{\kappa}$ is a random direction.
At each time step, the ion positions are checked to be within a cylinder of radius $r_0$ and length $L$ or are withdrawn from the simulation.
The performance of the code is limited by the Coulomb interaction evaluation, so the computation time scales as the square of the ion number. The double precision Coulomb force subroutine evaluates $5\times 10^7$ Coulomb terms per second on a 3~GHz CPU. Using multi-core CPU's, we observe a proportional speed-up.

\begin{figure}[pb]
\centerline{\includegraphics[width=3.6cm]{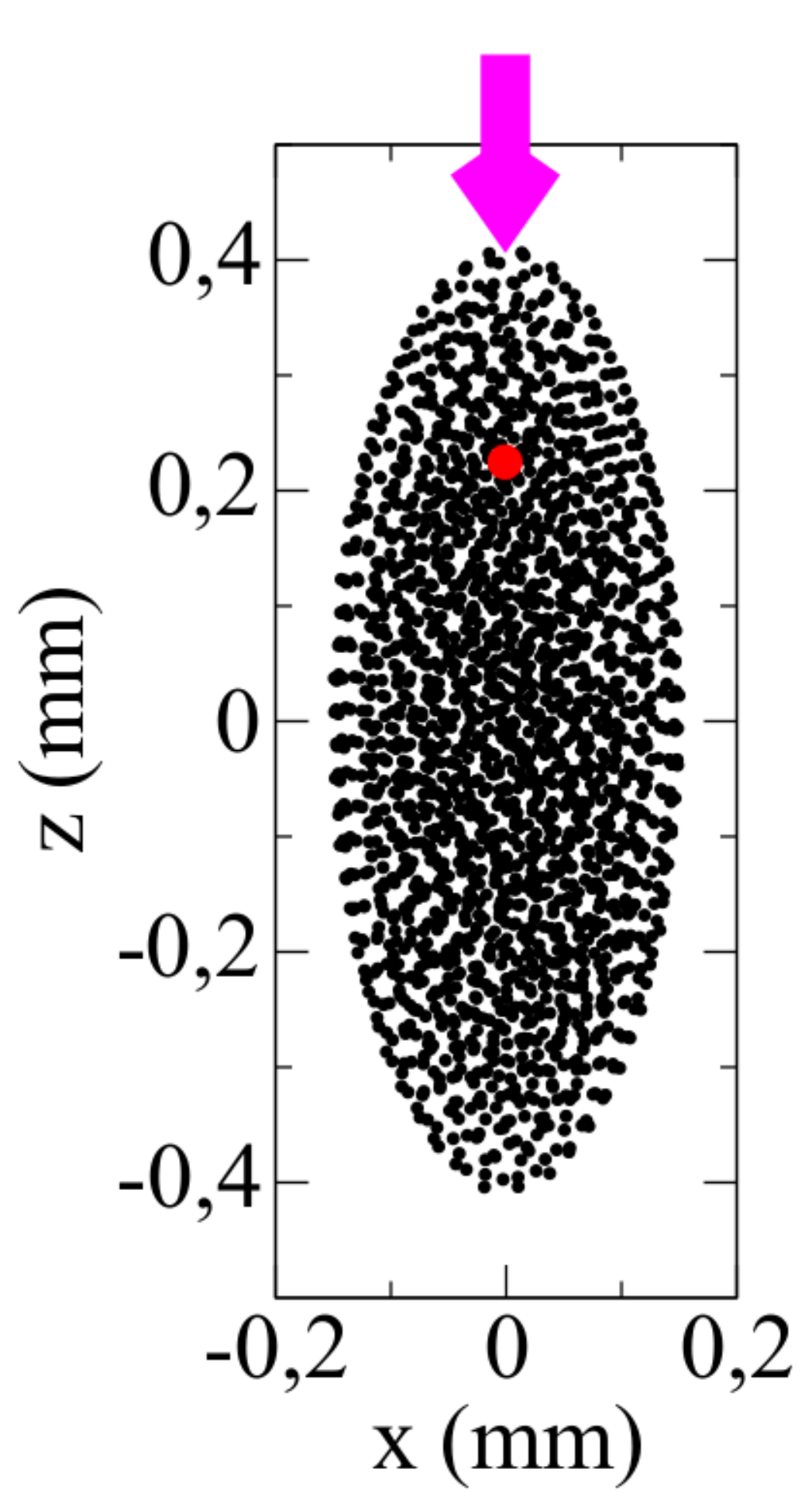}
\includegraphics[width=9cm]{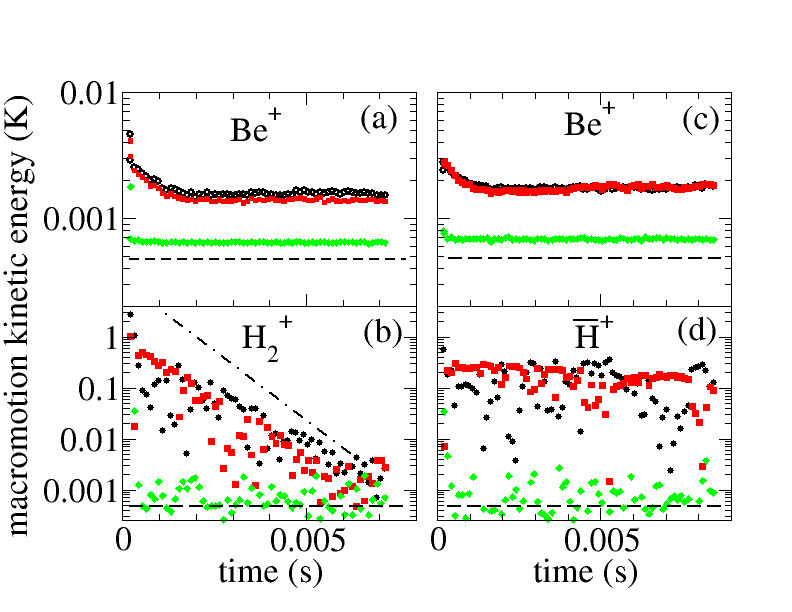}}
\caption{{\bf Left}: Snapshot of a 2000 Be$^+$ and 1 H$_2^+$ ion cloud. The sympathetically cooled H$_2^+$ ion is the red circle. The purple arrow shows the cooling laser direction. The cooling laser is aligned on the trap axis $z$ with a waist $w_0=100\ \mu$m located at the trap center. The laser detuning is $-\Gamma/2$ and the laser intensity on the axis is $I_{\rm sat}$.
{\bf Graphs}: Macromotion kinetic energy for each degree of freedom, averaged over 170 RF periods for 2000 laser cooled Be$^+$ ions (a,c) and for one H$_2^+$ (b) or $\bar{\rm H}^+$ (d) sympathetically cooled ion. Black circles: $x$, red squares: $y$, green diamonds: $z$. The horizontal dashed line corresponds to the Doppler cooling limit temperature $k_BT_D=\hbar\Gamma/2$ leading to $k_BT_D/2=3.2\times10^{-27}$~J. The dash-dotted line in (b) corresponds to a 1~ms exponential decay behavior.
For all the graphs, the trap parameters appearing in Eq.~(3) are: $r_0 = 3.5$~mm, $U_0 = 1$~V, and $\Omega = 2\pi \times 17$~MHz. The integration time step is $2 \times 10^{-10}$~s. (a) and (b): $V_{\rm RF}=356$~V, $\omega_{z}=500$~kHz for $m=1$. (c) and (d): $V_{\rm RF}=200$~V, $\omega_{z}=300$~kHz for $m=1$.}\label{fig_ion_cloud}
\end{figure}

A $N_{lc}$ laser cooled Be$^+$ ion cloud is numerically prepared and relaxed to equilibrium, and $N_{sc}$ $\bar{\rm H}^+$ or H$_2^+$ ions are introduced along the trap axis, next to the Be$^+$ ion cloud (see Fig.~\ref{fig_ion_cloud}) corresponding to a potential energy of a few meV.
One can distinguish two cooling phases. At the beginning of
the cooling process, the sympathetically cooled ion goes in and out the Be$^+$ ion cloud, and
only periodically interacts with the coolant ions, progressively losing secular kinetic energy.
During this first phase, the Be$^+$ ion cloud is not crystallized.

\begin{figure}[pb]
\centerline{
\includegraphics[width=6.5cm]{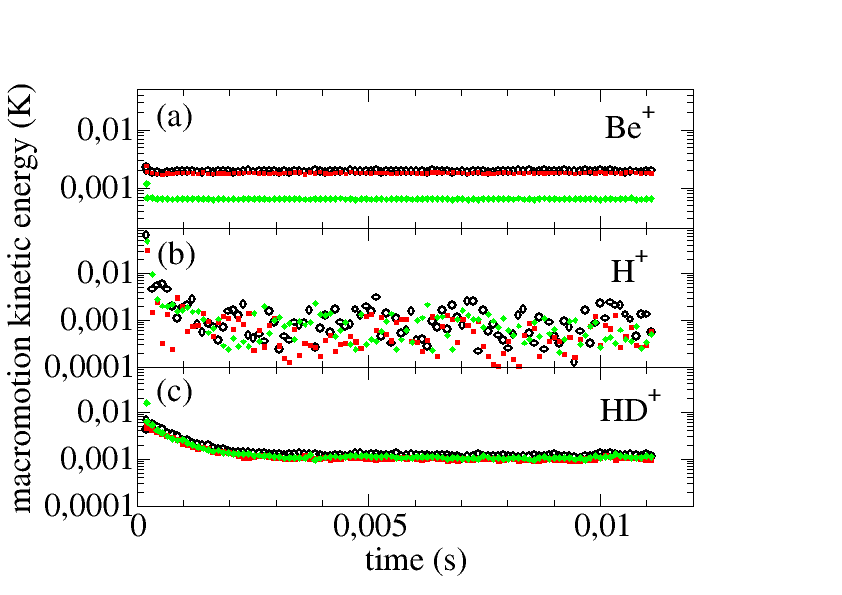}
\includegraphics[width=6.5cm]{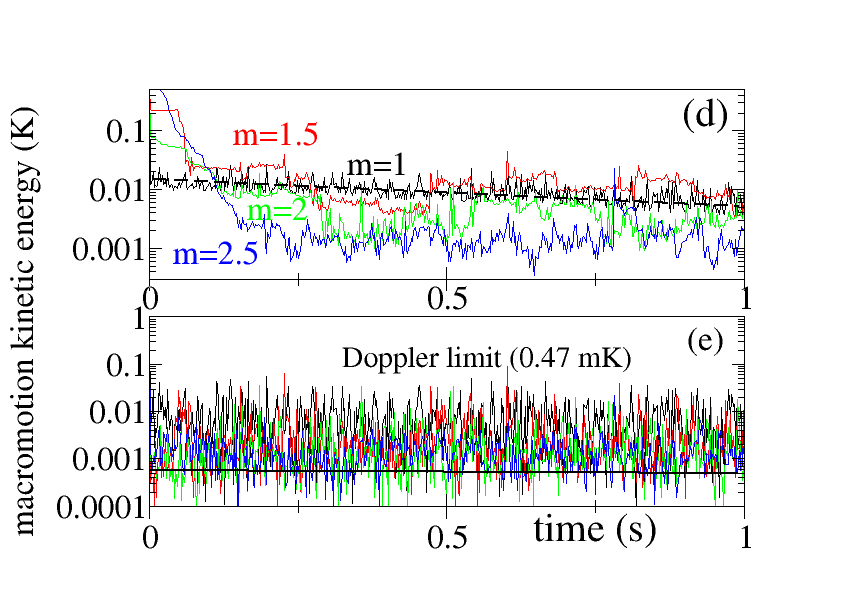}
}
\caption{{\bf Left}: Sympathetic cooling dynamics of a 1800 Be$^+$/ 200 HD$^+$/ 1 $\bar{\rm H}^+$ mixed ion cloud. Numerical parameters as in Fig.~\ref{fig_ion_cloud}(d). The $\bar{\rm H}^+$ initial position is on the trap axis, 0.5~mm from the center, corresponding to a 4.6~meV potential energy.
{\bf Right}: Doppler cooling dynamics of a Be$^+$/X$^+$ ion pair in the precision trap. The mass of the sympathetically cooled ion corresponds either to real (black $m=1$, green $m=2$) or fictitious ion masses ( blue $m=2.5$, red $m=1.5$). (d) 3D kinetic energy, (e) kinetic energy of the $z$ motion.}\label{fig_cooling_Be_HD_H}
\end{figure}

Once the sympathetically cooled ion gets embedded in the Be$^+$ cloud, a more efficient cooling phase then starts finally leading to a mixed species Coulomb crystal as shown in the left part of Fig.~\ref{fig_ion_cloud}, which illustrates the second phase of the sympathetic cooling dynamics for a single H$_2^+$ or $\bar{\rm H}^+$ ion by 2000 Be$^+$ laser-cooled ions.
We plot the averaged macromotion kinetic energy in the $x$, $y$ and $z$ directions (see figure~\ref{fig_ion_cloud} caption). In the case of H$_2^+$, we observe an exponential decay of the transverse kinetic energies down to the Doppler limit with a time constant of 1~ms, and a much faster decay of the axial kinetic energy, indicating the feasibility of fast sympathetic cooling for a 9/2 ion mass ratio.
For $\bar{\rm H}^+$, the situation is quite different. Whereas the axial motion is quickly damped to the Doppler limit, the competition between RF heating and sympathetic cooling in the transverse direction leads to a high transverse $\bar{\rm H}^+$ kinetic energy corresponding to temperatures in the K range. Indeed, the motional coupling between two particles of different masses rapidly decreases with the mass ratio.
It is thus important to work out more efficient sympathetic cooling schemes. One solution is to use an intermediate mass ion\cite{Zhang2007} such as HD$^+$ with a mass of 3. The left part of Fig.~\ref{fig_cooling_Be_HD_H} shows that starting with a Coulomb crystal made of 1800 Be$^+$ and 200 HD$^+$ ions, ms $\bar{\rm H}^+$ cooling times are achievable.

\subsection{Photodetachment constraints}\label{sec_photodetach}
The photodetachment threshold of $\bar{\rm H}^+$ (1.64~$\mu$m) is well below the 313~nm Be$^+$ cooling photon energy, so the cooling laser
beam may photodetach the excess positron.
For a Be$^+$ cooling beam at saturation intensity ($I_{sat}=2\pi^2\hbar c\Gamma/3\lambda^3$=0.82~mW/mm$^2$ with $\Gamma=2\pi\times 19.4$~MHz), the photon flux is $\Phi=7.9\times 10^{16}$~photon/s/cm$^2$.
The photodetachment cross section of H$^-$ at
313~nm\cite{Chandrasekhar1958,Smith1959} is $\sigma=2\times10^{-17}$~cm$^2$, leading to $\sigma \Phi=1.6$~s$^{-1}$. Under those conditions, the $\bar{\rm H}^+$
lifetime is less than 1s. It can be made longer using a lower cooling intensity or by using a quasi continuous cooling beam with a
reduced duty cycle. In a RF trap, the trapping effective potential is tighter for light ions than for heavy
ones.
Taking advantage of this fact, one might use a hollow laser beam in a Gauss-Laguerre mode\cite{Schmiegelow2012} L$_0^1$ to cool the Be$^+$
while the $\bar{\rm H}^+$ ions which are strongly confined very close to the trap axis are exposed to a negligible amount of laser radiation at 313nm.

\section{Ground state sympathetic cooling of a Be$^+$/$\bar{\rm H}^+$ ion pair in the precision trap}\label{sec_precision_trap}
As mentioned in the Introduction, the Doppler limit temperature is not low enough for the GBAR project, so the $\bar{\rm H}^+$ ion will be injected in a precision trap to form a Be$^+$/$\bar{\rm H}^+$ ion pair on which ground state Raman side band cooling can be performed~\cite{Leibfried2003}.

\begin{figure}[pb]
\center{
\includegraphics[width=4.7cm]{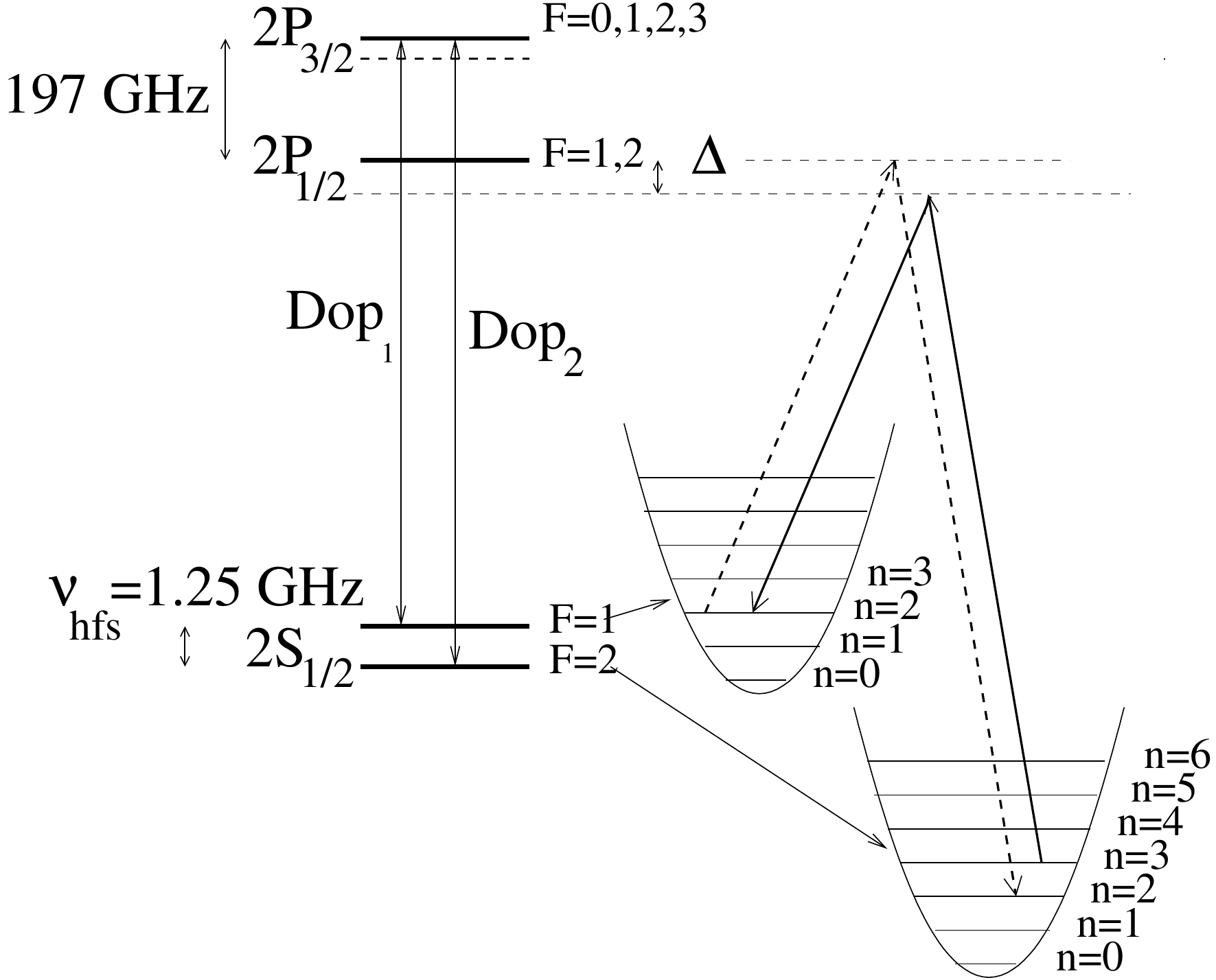}
\includegraphics[width=6cm]{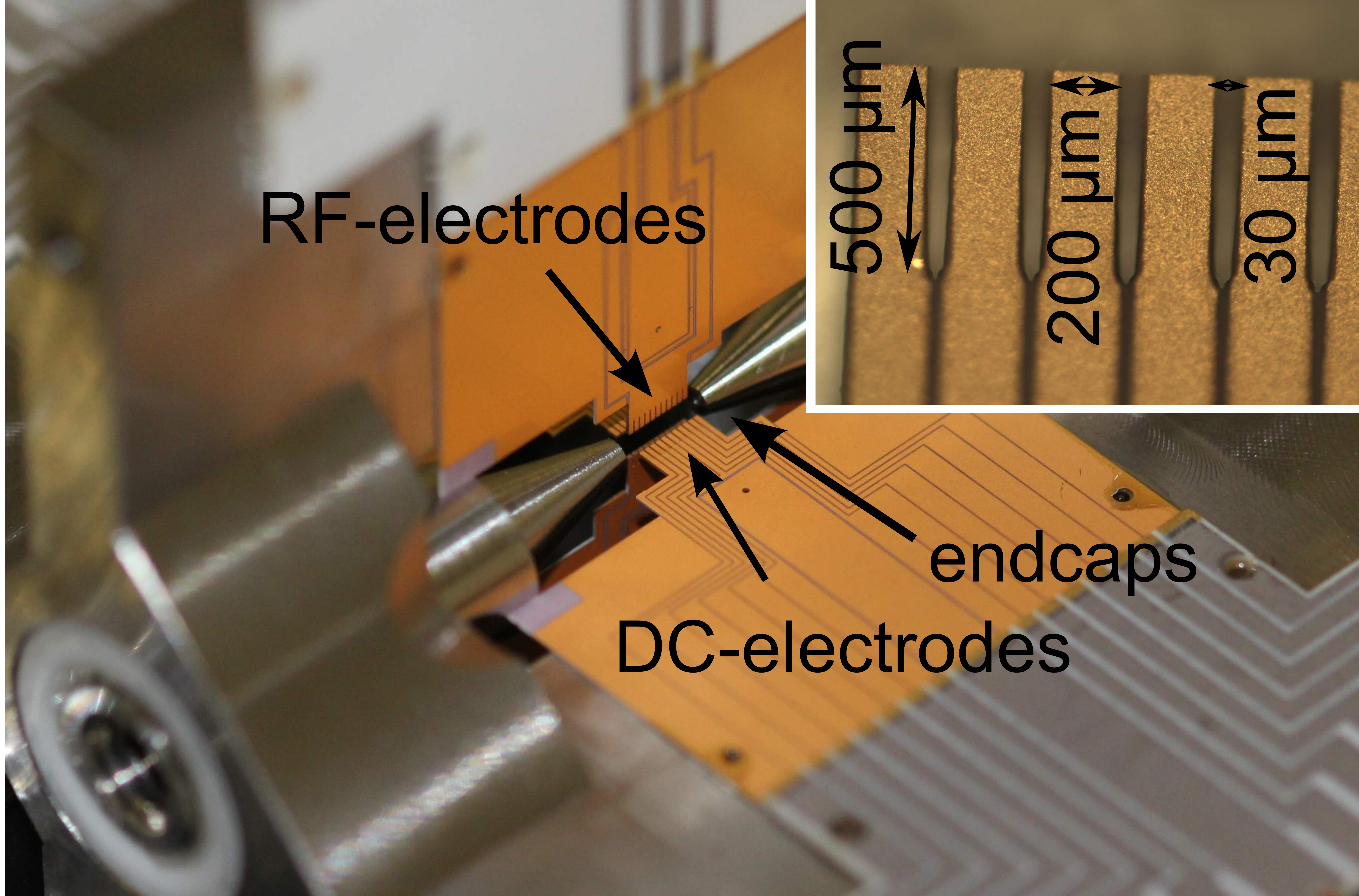}}
\caption{{\bf Left}: Energy levels of a Be$^+$/$\bar{\rm H}^+$ ion pair including the ground and excited level of Be$^+$ and showing the harmonic ladder of one of the motional eigenmodes. The thin solid arrows show the Doppler cooling and repumping transitions. The thick solid arrows show the stimulated Raman transition, and the dashed arrows the spontaneous Raman transition. {\bf Right}: Miniaturized trap of similar geometry to the precision trap to be used in the GBAR experiment.}\label{fig_raman_transition_precision_trap}
\end{figure}

The precision trap (see Fig~\ref{fig_raman_transition_precision_trap})\cite{Schnitzler2009} consists of four gold coated, micro-fabricated alumina chips which are arranged in an x-shaped configuration and two endcaps made from titanium. The endcaps are pierced with a hole with a diameter of 600~$\mu$m to enable ion injection into the trap. Two of the chips provide the RF-field and two the DC trapping potential respectively. The chips have 11 electrodes each to shape  the axial potential what for the voltages can be controlled with a custom built digital-to-analog converter with a voltage resolution of 300~$\mu$V and a time resolution of 400~ns. The distance between the chips is 960~$\mu$m.
The trap is driven by an RF-voltage with a frequency $\Omega= 2\pi\times 56 \text{ MHz}$ and a peak-to-peak amplitude $V_{RF}= 176\text{ V}$. This leads to a q parameter for $Be^+$ of 0.05 and for $\bar{H}^+$ of 0.45. The center DC-electrode is held at -1.5~V to provide axial confinement. This voltage configuration leads to a axial and radial secular frequencies $\omega_z=2\pi\times 1.9 \text{ MHz}$ and $\omega_{x,y}=2\pi\times 8.7 \text{ MHz}$ for $\bar{\rm H}^+$.
The corresponding ground state kinetic energies are 0.09 and 0.41~mK.

After extraction from the capture trap, a cold $\bar{\rm H}^+$ ion is injected into the precision trap through the end cap. The right part of Fig.~\ref{fig_cooling_Be_HD_H} shows sympathetic Doppler cooling of a Be$^+$/X$^+$ ion pair where the precision trap is modeled as an ideal linear Paul trap. The cooling time strongly depends on the X$^+$ mass and can be larger than seconds in the case of $\bar{\rm H}^+$. Again, the z-motion cooling is much faster due to the absence of RF heating.

In the precision trap, the cold Be$^+$ and $\bar{\rm H}^+$ ions are coupled harmonic oscillators.
The Doppler limit temperature (0.47~mK) corresponds to excitations of a few quanta that can be further laser cooled.
The motional couplings of an ion pair confined in a RF Paul trap have been evaluated by W\"ubbena et al.\cite{Wubbena2012}.
In the $x$, $y$ or $z$ directions, the individual ion trajectories can be expanded on {\it in-phase}
and {\it out-of-phase} eigenmodes as
\begin{eqnarray}
u_{1}(t)&=&b_1 z_{in}\sin(\omega_{in}t+\phi_{in})+b_2 z_{out}\sin(\omega_{out}t+\phi_{out})\\
u_{2}(t)&=&\sqrt{\frac{m_1}{m_2}}b_2 z_{in}\sin(\omega_{in}t+\phi_{in})-\sqrt{\frac{m_1}{m_2}}b_1 z_{out}\sin(\omega_{out}t+\phi_{out}),
\end{eqnarray}
where the amplitudes $z_{in}$ and $z_{out}$ and phases $\phi_{in}$ and $\phi_{out}$ depend on the initial conditions. The motional coupling coefficients $b_1$ and $b_2$ depend on the particle masses and trapping conditions and verify $b_1^2+b_2^2=1$ (see Eq. (14) and (17) in\cite{Wubbena2012}). For the axial motion, we get $b_{1,z}=0.982$ and $b_{2,z}=0.187$. For the transverse motion, assuming $\omega_{x,y}=1.1\ \omega_z$, we get $b_{1,x,y}=0.99971$ and $b_{2,x,y}=0.017$, which can be compared to $1/\sqrt{2}\approx0.707$ in the case of an ion pair with equal masses, or 1 for a single ion.
Figure~\ref{fig_raman_transition_precision_trap} shows the Be$^+$ electronic energy levels with the confined ion vibrational structure.
Raman side band cooling consists in using an off-resonance stimulated Raman transition and a resonant spontaneous Raman transition to decrease the vibration number down to 0. A stimulated Raman transition is a coherent process. The time required to drive a $\pi$-pulse is inversely proportional to the dipole matrix element, i.e. to the coupling coefficient $b_2$. For a Be$^+$/$\bar{\rm H}^+$ ion pair, it is at most 60~times longer than for a single ion. In the latter case, Raman sideband cooling to the ground state for the three degree of freedom can be performed within a few tens of ms\cite{Rosenband2007,Chou2010}. For a Be$^+$/$\bar{\rm H}^+$ ion pair, it may thus be achieved for the 6 degrees of freedom within 1~s.

Eigenmodes with frequencies in the 2-8~MHz range make it possible to efficiently laser cool the ion pair to
its vibrational ground state. The corresponding velocity dispersion is still slightly too large for the GBAR experiment.
We thus propose to adiabatically ramp down the trapping stiffness down to $\approx$~30~kHz radial and longitudinal oscillation frequencies by slowly lowering the trapping voltage within about 500~ms.
30~kHz oscillation frequencies correspond to $\Delta v\approx$~8~cm/s velocity dispersions which amply satisfies the GBAR requirements.

\section{Conclusion and perspectives}
We have discussed the challenges and feasibility of antihydrogen positive ion $\bar{\rm H}^+$ sympathetic cooling down to $\mu$K temperatures for the GBAR project. We here summarize the results.
\begin{itemize}
\item Capture efficiency of a 1~keV $\bar{\rm H}^+$ ion bunch in a biased linear trap can be larger than 50\% with a 25~eV kinetic energy dispersion and a trap depth of 20~eV. The efficient capture of an ion bunch with 300~eV dispersion requires out-of-reach trapping conditions, indicating that the antiprotons delivered by the ELENA ring have to be cooled beforehand to decrease the energy dispersion by at least a factor of 10.
\item Sympathetic Doppler cooling by laser cooled Be$^+$ ions is shown to be efficient if the ions remain embedded in the Be$^+$ ion cloud.  This means that one has to use large Be$^+$ ion clouds filling the capture trap. The numerical simulations show that the cooling efficiency is much better with a 9/2 rather than with a 9/1 mass ratio. In the latter case, the efficiency is dramatically improved using a third species of mass 3, i.e. HD$^+$ ions. A possible experimental scheme is then to prepare a laser cooled Be$^+$ ion cloud and to inject HD$^+$ ions from an external ion source (in order not to increase the pressure in the vacuum chamber) before the $\bar{\rm H}^+$ bunch intake. Sympathetic Doppler cooling of energetic $\bar{\rm H}^+$ is the most challenging step and is still an open problem, which has to be tackled both experimentally and using numerical simulation. To that end and in order to perform numerical simulations of the cooling dynamics with large number of ions ($>$ 10000), the code will be implemented on massively parallel Graphic Processing Units (GPU). From the experimental point of view, this step will be first tested using matter ions H$_2^+$ and H$^+$ (protons).
\item The transfer of a single $\bar{\rm H}^+$ ion from the capture trap to the precision trap, which was not discussed here, will be done using standard ion beam optics for injection through the drilled end cap. The experimental protocol will be worked out first with matter ions with Ca$^+$/Be$^+$ and then Be$^+$/H$_2^+$ and Be$^+$/protons before being implemented on GBAR. Here, the main issue is to avoid heating the ion during the transfer to secure a fast re-capture and sympathetic Doppler cooling of the ion pair.
\item We have shown that once a Be$^+$/$\bar{\rm H}^+$ ion pair is prepared at the Doppler limit temperature, Raman sideband sympathetic cooling down to the vibrational ground state of the trap is feasible with less than one second, preparing a $\bar{\rm H}^+$ ion with a velocity dispersion of about 1~m/s.
\item The velocity dispersion can be decreased to about 10~cm/s by adiabatically ramping down the trapping stiffness by a factor of 100, within less than 0.5~s. At that point, the velocity dispersion of the antihydrogen produced by threshold photodetachment of the excess positron is dominated by the recoil due to the e$^+$ ejection.
\end{itemize}

\section*{Acknowledgments}
We thank Jofre Pedregosa for helpul discussions and Dominique Delande for introducing us to OPEN MP parallelization. We also thank the COST action MP1001-IOTA, NanoK-Ifraf Resima grant and ANR/DFG ANR-13-IS04-0002-01 BESCOOL grant.

\end{document}